\documentstyle[11pt,paspconf]{article}

\begin{document}

\title{Galactic Structure From Infrared Surveys}
\author{Andrew Gould\altaffilmark{1}}
\affil{Dept of Astronomy, 174 W. 18th Ave., Columbus, OH 43210-1106}

\altaffiltext{1}{Alfred P.\ Sloan Foundation Fellow}

\begin{abstract}

By combining the 2MASS and DENIS infrared surveys with the USNO-B 
proper-motion catalog, it will be possible to map the structure of the Galaxy 
in unprecedented detail.  The key parameter that these surveys measure 
together, but neither measures separately, is $t_*=r_*/v_\perp$, the time it 
takes a star moving at transverse speed $v_\perp$ to cross its own radius 
$r_*$.  This parameter cleanly separates white dwarfs, main sequence stars, and
giants, and simultaneously separates disk from spheroid stars.  The infrared 
photometry then yields photometric parallaxes for the individual stars, while 
the proper motions give kinematic information.  Hence, it becomes possible to 
measure the physical and velocity distributions of the known Galactic 
components and to identify new ones.  The analysis of such a huge data set 
poses a major technical challenge.  I offer some initial ideas on how to meet 
this challenge.

\end{abstract}

\keywords{galactic structure, surveys, astrometry}

\section{Introduction}

        50 years from now, our catalogs will probably list position, color, and
velocity information for $10^{10}$ stars, a large fraction of all the stars
in the Galaxy and about 5 orders of magnitude more than today.  But it is not
clear how much more we will know about Galactic structure than we do today.
For example, how much more would we know about the air in this room if we
catalogued the position and velocity of its $10^{34}$ molecules, than we do
just from the temperature, pressure, and composition?  It is appropriate
to contemplate the construction of such a gigantic star catalogue because
within 2 or 3 years we will be half way there (logarithmically).  It is not
at all obvious how we can make use of the forthcoming vast quantities of data 
to learn about the structure of our Galaxy.

        Traditionally, the Galaxy is broken down into components,
and as more data are acquired, more components are added: disk, 
thick disk, spheroid, bulge/bar, spiral arms (e.g. Bahcall 1986).  
Each component is described
by a few parameters, e.g., scale height, scale length, and
luminosity function
for the disk.  The study of Galactic structure is basically one of identifying
these components and measuring their parameters with increasing precision.

        We might at first imagine taking all of our new data on the one hand,
and all the global parameters describing the various Galactic components on
the other, and dumping them both into a huge maximum likelihood program that
will in turn spit out a detailed description of the Galaxy.  
However, such an approach would overwhelm even the most optimistic estimate
of near-term computing advances.

\section{Individual-Star Parameter Counting}

        From the standpoint of Galactic structure, the three most important
quantities to learn about each star are its distance, $d$, and its type,
characterized by its radius, $r_*$, and its temperature, $T$.  In order to
learn anything about these quantities from photometry, one must simultaneously
measure the extinction, $A_V$.  The two infrared surveys, 2MASS and DENIS, will
measure fluxes in only three bands, $JHK$ and $IJK$ respectively.  Simple
parameter counting tells us we cannot determine four quantities from three
observables.  In fact, we can extract only $A_V$, $T$, and $\theta_*=r_*/d$, 
the
angular radius of the star.  The deviation of the two observed colors
from the black-body two-color relation yields $A_V$, and the dereddened color
gives $T$.  From the dereddened color and flux plus the Planck law one 
then knows $\theta_*$.  A white dwarf at 1 pc, a subdwarf at 25 pc,
a disk dwarf at 40 pc, and a disk giant at 500 pc will all look approximately
the same to an infrared survey if they have the same temperature.  One way 
out might be to get a fourth color, but in fact since stars are approximately
black bodies, the fourth color can already be predicted from the other three.

        Thus, even to isolate the individual Galactic components, let alone
measure them, requires more information than we are going to get from an
infrared survey.  Fortunately, before these surveys are completed, the
USNO-B astrometric catalog should be available.  

\section{Proper Motions}

        From this catalog (combined
with an additional epoch from 2MASS/DENIS), one should be able to
achieve an astrometric precision of
\begin{equation}
\sigma_\mu \sim 3\, {\rm mas}\,{\rm yr^{-1}} \sim 14\,{\rm km}\,
{\rm s}^{-1}\,{\rm kpc}^{-1}\qquad \rm (North),
\label{eqn:sigmamu}
\end{equation}
and perhaps a factor 2 worse in the South (D.\ Monet 1998, private 
communication)

        Photometry plus proper motions together yield a very curious quantity,
\begin{equation}
t_*\equiv {r_*\over v_\perp} =  {r_*/d\over v_\perp/d}={\theta_*\over \mu},
\label{eqn:tstar}
\end{equation}
the time it takes a star to cross its own radius.  Why is this quantity of
any interest?  Because, as Table \ref{tab:one} shows, it separates out very 
nicely different populations of stars.

\begin{table}
\caption{Stellar Radius Crossing Times}\label{tab:one}
\begin{center}\normalsize
\begin{tabular}{lrrr}
Star Type & $r/r_\odot$ & $v_\perp/({\rm km}\,{\rm s}^{-1})$ & $t_*/\rm min$ \\
\tableline
&&&\\
Halo White Dwarf &0.02 &200 &1\\
Disk White Dwarf &0.02 &40 &6\\
Halo Sub-Dwarf &0.5 &200 &30\\
Disk Dwarf &0.8 &40 &230\\
Halo Gaint &10 &200 &600\\
Disk Gaint &10 &40 &3000\\
\tableline
\end{tabular}
\end{center}
\end{table}

Of course, any individual star will still be ambiguous at some level.  However,
in an analysis of millions of stars, all that is important is that
most of the stars in a given component are so identified with $\ga 70\%$
probability.  Maximum likelihood can then easily sort out the components
statistically.  Since individual components are separated in $t_*$
by a factor 3 to 5, this condition is well satisfied.

        Virtually all of the stars detected by 2MASS have counterparts
on the POSS E ($R$ band) plates in the North or the equivalent ESO plates
in the South, so proper motions should be available for all of them.

\section{The Near Zone}

        At 1 kpc, a disk star has a proper motion $\mu\sim 8\,{\rm mas}\,
{\rm yr}$ which should be detectable at least in the North.  Main-sequence
disk stars down to mid-G  $(M_K<4)$
should be detectable in $K$ at the $10\,\sigma$ level at this distance.
There are about $10^7$ such stars.  There are also about $10^6$ detectable
thick-disk star and $10^4$ detectable halo subdwarfs.  The proper motions
of these latter two classes of stars would be detectable at much larger 
distances (2 kpc and 5 kpc respectively).  However, the fraction of 
main-sequence stars that are detectable photometrically
drops rapidly past 1 kpc.  
Nevertheless, of order $10^4$ halo giants will be detectable in both the
photometric and proper motion surveys.

        What sort of questions can be asked about the stars in this zone?
First, of course, one can measure the parameters of the known components
much more precisely than they have been measured before.  For example, the
disk is usually parameterized by a scale height $h\sim 200$ pc, and a scale 
length of $H\sim 3$ kpc, while the thick disk has $h\sim 600$ pc and 
$H\sim 3$ kpc (Gould, Bahcall, \& Flynn 1997).  The spheroid is parameterized
by a power-law density fall-off $\nu\sim -3.1$ and a flattening $c/a=0.8$
(Gould, Flynn, \& Bahcall 1998).  

        However, with the vast new quantities of data that will soon be
available, we should be able to ask many other questions about Galactic
structure.  What are the velocity distributions of the various components?
Traditionally, one has tried to measure only the first and second moments,
but higher moments will also be accessible.  Gould \& Popowski (1998) found,
for example, that the kurtoses of the 
halo as determined from $\sim 150$ RR Lyrae stars are 
$\sim 2$, 3 and 4
in the three principal directions, compared to 3 for a Gaussian and 1.8
for a sharp-edged box distribution.  This is undoubtedly telling us something
about the origin of the stellar halo, but what?  Much more precise and
detailed information about all three components will be obtained from the
IR and proper motion surveys.  Are the scale height and scale length of the
disk and thick disk functions of spectral type?  Are there additional
components?  For example, Sommer-Larsen \& Zhen (1990) and 
Gould et al.\ (1998), each some found evidence for a two-component halo, one
roughly round and the other highly flattened (but not rotating).  
The high vertical
kurtosis found by Gould \& Popowski (1998) could also be regarded as evidence
for this.  Or it is possible, that our current disk/thick disk dichotomy should
be replaced by a continuum of components, as suggested by Norris \& Ryan 
(1991).  

\section{Complementarity of SDSS}

     If there are a continuum of components, they will be difficult to
distinguish from discrete components using IR-survey plus proper-motion
data alone.  The argument codified in Table \ref{tab:one} was that
discrete components could be separated based on their radius crossing
times, $t_*$.  However, this assumed that the components are separated by a 
factor two in $t_*$.

     To distinguish finer graduations requires a metallicity-sensitive
indicator.  This will not be available from 2MASS and DENIS, but will
be from the Sloan Digital Sky Survey (SDSS), which has 5-color photometry
out to the atmospheric cutoff.  Unfortunately, in its present
incarnation, SDSS will saturate at 15th magnitude, implying very little
overlap with $K<14$ surveys (and then only for stars with poor IR photometry.)
Moreover, SDSS will cover only regions far from the plane.

     However, from the standpoint of the statistical separation of components,
it is not necessary that there be SDSS and IR photometry for the same stars.
It is only necessary that the SDSS stars be close enough that they have
proper motions reliable to better than a factor 2.  The transverse motions of
thick-disk stars are $\sim 100\,\rm km\,s^{-1}$, implying factor 2 
proper motions errors at a distance $\sim 3\,$kpc.  At this distance, the
entire main sequence is unsaturated.

\section{Mass Distribution of Disk}

	Traditionally, measurements of the local mass density have relied on
stellar radial velocity and density measurements in a cone perpendicular
to the Galactic plane.  
(e.g. Bahcall 1984; Kuijken \& Gilmore 1989, 1991; Bahcall, Flynn \& Gould 
1992;
Flynn \& Fuchs 1994).  Cr\'ez\'e
et al.\ (1998) pioneered a radically different type of survey based on
Hipparcos parallaxes and
proper motions of $\sim 3000$ nearby stars.  This study was by far
the most sensitive to mass close to the Galactic plane.	 They found
that the local mass density is almost completely accounted for by visible
material.  However, because of the Hipparcos magnitude limit $(V<8)$ they
were able to probe only within 100 pc of the Sun.  Moreover, since they
were mainly restricted to bright (and hence young) A and F stars, it is 
possible that their sample was not dynamically mixed and hence subject to
systematic errors.

	IR surveys (plus proper motions) will provide a much more robust
measurement of the density close to the plane, first because the sample
will be three orders of magnitude larger than the Hipparcos sample, and
second because the stars will be primarily G and later and thus older and
more dynamically mixed and so less subject to systematic errors.

	Of course, unlike the Hipparcos stars, these stars will not have
trigonometric parallaxes.  One will be forced to rely on photometric 
parallaxes.  Will the inevitable distance errors, including systematic errors,
compromise the result.  Remarkably, no.  The density, $\rho$, 
is given by Poisson's equation
\begin{equation}
\rho = -{1\over 4\pi G}\,{d K_z\over d z},
\label{eqn:poisson}
\end{equation}
where $K_z(z)$ is the vertical acceleration as a function of height $z$ above
the plane.  For simplicity, I consider an isothermal population of stars,
but the argument I am about to give applies equally well to any velocity
distribution.  Then
\begin{equation}
K_z = -\overline{v_z^2}\ {d\ln \nu\over d z},
\label{eqn:kz}
\end{equation}
where $\overline{v_z^2}$ and $\nu(z)$ are the vertical velocity dispersion and 
vertical density profile
of some stellar population.  Now, suppose that there is a
survey close to the plane (e.g. $|b|<10^\circ$), and suppose that
all photometric
distances are systematically overestimated by a factor $\alpha$ relative to
the true distances.  Then  $z\rightarrow \alpha z$ will be overestimated
from the measured angular position, and $v_z\rightarrow \alpha v_z$ will
be overestimated from the measured proper motion.  
($\nu\rightarrow \alpha^{-3}\nu$ will be underestimated, but this has no
effect, because only the derivative of $\ln\nu$ enters eq.\ \ref{eqn:kz}).  
Hence, from equation
(\ref{eqn:kz}) $K_z$ will be overestimated by $\alpha^2/\alpha=\alpha$, so
that from equation (\ref{eqn:poisson}), $\rho$ will be properly estimated
despite the errors.  Of course, $v_z$ cannot be directly estimated from
the proper motion and the distance.  Rather 
$v_z= \mu_b d\sec b+v_\parallel\tan b$, where $v_\parallel$ is the
component of motion parallel to the Galactic plane.  That is,
$\overline{v_z^2}= \overline{(\mu_b d)^2}\sec^2 b-
\overline{v_\parallel^2}\tan^2 b$.
To obtain $\overline{v_z^2}$, one must therefore independently estimate
$\overline{v_\parallel^2}$ and subtract it.  For $|b|<10^\circ$, $\tan^2 
b\la 0.03$, so the
systematic error in this correction is negligible.  It should be possible to
obtain excellent results for $z<170\,$ pc (assuming sensitivity to stars
at 1 kpc).  For higher $b$, the systematic errors will become progressively
worse.  However, it will be possible to map the mass density with much
greater precision than previously, especially close to the plane.

\section{The Far Zone}

	Paczy\'nski \& Stanek (1998) have recently suggested $I$-band 
luminosities of
clump giants (metal rich core-helium burning stars) as standard candles.
Whether clump giants eventually turn out to have the same $I$-band 
luminosities independent of age, metallicity, and other environmental
factors remains open to question.  However, the bulge clump giants can
reasonably be expected to have a common set of environmental characteristics
and therefore act at least as {\it relative} standard candles for the bulge.
That is, they may or may not tell us about the galactocentric distance, but
they can certainly trace the density structure of the bulge.  
With $K_0\sim 13$ and $I_0\sim 14$, they will be visible in the bulge in all 
regions with $A_V<9$.  They can be dereddened in the standard way.  Hence,
they will provide a three dimensional map of the bulge.  The typical
proper motions (from internal dispersion) are only $\sim 3\,\rm mas\,yr^{-1}$.
Since these are the same order as the errors, it may be possible to obtain
statistical information on the velocity dispersion as a function of position,
although systematics will be a major concern.

	Spiral arms contain large numbers of early type stars, probably
of common metallicity, which should provide good relative distances.  However,
they will yield almost no proper-motion information.

	It will also be possible to map the three dimensional distribution of
dust using early type stars and clump giants.

\section{Likelihood}

	We can imagine, then, maximizing the likelihood of a given model
represented by say 100 parameters, given the observation of say 400,000,000
data values.  The parameters being disk scale height, scale length, luminosity
function, thick disk parameters, spheroid parameters, velocity ellipsoids,
kurtoses, bar parameters, spiral arm densities and pitch angles, etc.  The
data are $JHK$ photometry and proper motions for the stars, plus other
colors when available from various sources.  In principle, one could just
dump all this in a giant black-box likelihood program and wait for it to
spit out the answer.  A big job, perhaps, but one can always hope that
computers will improve sufficiently by then.  Good luck!

	In fact, new computational methods will be required.  The evaluation
of the likelihood function requires the prediction of the probability of
hundreds of millions of observations for a huge ensemble of models spanning
a vast parameter space.  Here I outline one possible approach based on the
one used by Gould et al.\ (1998).

	Consider a bin in observation space 
$J, H, K, \alpha, \delta,\mu_\alpha,\mu_\delta$.  The probability of observing
$n$ stars in this bin is

\begin{equation}
P_{J, H, K, \alpha, \delta,\mu_\alpha,\mu_\delta} =
{1\over n!}\exp(-\tau_{J, H, K, \alpha, \delta,\mu_\alpha,\mu_\delta})
\tau_{J, H, K, \alpha, \delta,\mu_\alpha,\mu_\delta}^n
\label{eqn:pdef}
\end{equation}
where $\tau$ is the expected density of stars times the volume of the
parameter-space bin.  Then, by definition, $L=\prod_\beta P_\beta$ where
$\beta=({J, H, K, \alpha, \delta,\mu_\alpha,\mu_\delta})$.  Hence, 
\begin{equation}
\ln L = \sum_\beta n_\beta\ln\tau_\beta - \sum_\beta\tau_\beta - 
\sum_\beta\ln n_\beta!.
\label{eqn:lnl}
\end{equation}
In the limit of very small bins $n\rightarrow 0,1$, which implies 
that $\ln n_\beta! =0$.
Hence,
\begin{equation}
\ln L = \sum_{\beta,\rm detections} \ln \tau_\beta - N_{\rm exp}
\label{eqn:lnltwo}
\end{equation}
where $N_{\rm exp}$ is the total number of detections predicted by the model.

We can now write $\tau_\beta$ as a vector product of two types of quantities, 
one
of which depends on the choice of model parameters, and the other of which
is independent.  For example, assuming disk model parameters scale height $h$,
scale length $H$, and $K$-band luminosity function with 1-mag averaged bin
densities $\Phi_l$, one can write,

\begin{equation}
\tau_{i,j,k,p} = \sum_{l,m}\Phi_l \nu_{k,m}(h,H)
{\rm cmd}_{i(m),j(m),k,l,m} {\rm pm}_{k,p,m}(v_1,v_2,v_3,\sigma_1,\sigma_2,
\sigma_3).
\label{eqn:taueval}
\end{equation}
Here $i,j,k,p,q$ are observational bins, $i=(H-K)_0$, $j=K_0$, 
$k=(\alpha,\delta)$, and $p=(\mu_\alpha,\mu_\delta)$.  The model-dependent
density is given by $\nu_{k,m}(h,H)$ as a function of distance modulus $m$,
position vector $k$, and model parameters $h$ and $H$.  The model-dependent
observed proper motion distribution, ${\rm pm}_{k,p,m}$,
 is given as a function of $k$, $m$, and
proper-motion vector $p$, together with the model parameters $v_1,v_2,v_3$
bulk velocity and $\sigma_1,\sigma_2,\sigma_3$, dispersions.  However,
the color-mag diagram, which requires a cumbersome numerical integration
over observational errors, is given once and for all.

Similarly, one can write $N_{\rm exp}$ as
\begin{equation}
N_{\rm exp} = \sum_{k,l,m}\phi_l \nu_{k, m}(h,H){\rm cmd}_{{\rm tot},k,l,m},
\label{eqn:nexp}
\end{equation}
where
\begin{equation}
{\rm cmd}_{{\rm tot},k,l,m} = \sum_{i,j} {\rm cmd}_{i,j,k,l,m}.
\label{eqn:cmdtot}
\end{equation}
Again ${\rm cmd}_{{\rm tot},k,l,m}$, which involves an enormous numerical
integration, needs to be calculated only once.  Equation (\ref{eqn:cmdtot})
(and its derivatives with respect to all parameters) can be calculated
quite easily.  If one were to really try to evaluate equation 
(\ref{eqn:taueval}) for each of $10^7$ stars for each iteration, it would
still be a cumbersome project.  Just storing the matrix elements for
the model-independent function cmd$_{i,j,k,l,m}$ might prove unwieldy.
However, likelihood space can be searched on progressively larger subsets
of the data, say first using $10^5$ stars to locate the approximate solution,
and then increasing the sample to get more precise results.  The last
iterations would be time consuming, but there would be few of them.

\section{Conclusions}

	The completion of the 2MASS and DENIS surveys, together with the
USNO-B astrometric catalog will revolutionize our knowledge of Galactic
structure.  The astrometric input is required not only to obtain kinematic
information, but also to separate out the various components of the Galaxy,
which otherwise would be almost totally degenerate.  The computational
facilities required to interpret these data sets appear at first sight
daunting.  I have tried to give some initial suggestions on how the
analysis can be simplified.

\acknowledgments

I thank Martin Weinberg for several lengthy discussions that
were critical to the preparation of this paper.  
This work was supported in part by grant AST 97-27520 from the NSF and in 
part by grant NAG5-3111 from NASA.

\end{document}